\documentclass[iop]{emulateapj}
\pdfoutput=1
\usepackage{natbib}
\usepackage{threeparttable}

\newcommand{\HI}{\ensuremath{\mbox{\rm \ion{H}{1}}}}
\newcommand{\HII}{\ensuremath{\mbox{\rm \ion{H}{2}}}}
\renewcommand{\t}[1]{\mathrm{#1}}
\renewcommand{\H}{\ensuremath{\mathrm{H}}}
\newcommand{\htwo}{\ensuremath{\mbox{H$_2$}}}
\newcommand{\msun}{\ensuremath{M_\odot}}

\newcommand{\kms}{\mbox{km~s$^{-1}$}}
\newcommand{\xco}{\ensuremath{X_{\mathrm{CO}}}}

\newcommand{\ico}{\ensuremath{I_{\mathrm{CO}}}}
\newcommand{\cm}{\mbox{cm$^{-2}$}}
\newcommand{\e}{\ensuremath{\mathrm{e}}}
\newcommand{\avg}[1]{\mbox{$\langle{#1}\rangle$}}

\newcommand{\co}[1]{\mbox{$^{#1}$CO}}

\newcommand{\av}{\ensuremath{\mbox{$A_{\rm V}$}}}
\newcommand{\xunits}{\mbox{cm$^{-2}$ (K km s$^{-1}$)$^{-1}$}}
\newcommand{\counits}{\mbox{K km s$^{-1}$}}

\newcommand{\arc}{\mbox{$^{\prime\prime}$}}

\newcommand{\mad}{G216-2.5}

\shortauthors{Imara}

\begin{document}


\title{Rethinking a Mysterious Molecular Cloud}


\author{N. Imara}
\affil{Harvard-Smithsonian Center for Astrophysics,
    Cambridge, MA 02138}

\email{nimara@cfa.harvard.edu}

\begin{abstract}
I present high-resolution column density maps of two molecular clouds having strikingly different star formation rates.
To better understand the unusual, massive \mad, a molecular cloud with no massive star formation, the distribution of its molecular gas is compared to that of the Rosette Molecular Cloud.
 Far-infrared data from \emph{Herschel} are used to derive $N(\htwo)$ maps of each cloud and are combined with \ico~data to determine the CO-to-\htwo~ratio, \xco.   In addition, the probability distribution functions (PDFs) and cumulative mass fractions of the clouds are compared.
 For \mad, $\avg{N(\htwo)}=7.8\times 10^{20}~\cm$ and $\avg{\xco}=2.2\times 10^{20}$ \xunits; for the Rosette,  $\avg{N(\htwo)}=1.8\times 10^{21}~\cm$ and  $\avg{\xco}=2.8\times 10^{20}$ \xunits.  The PDFs of both clouds are log-normal for extinctions below $\sim 2$ mag and both show departures from log-normality at high extinctions.  Although it is the less-massive cloud, the Rosette has a higher fraction of its mass in the form of dense gas and contains $1389\msun$ of gas above the so-called extinction threshold for star formation, $\av = 7.3$ mag.  The \mad~cloud has $874\msun$ of dense gas above this threshold.
\end{abstract}



\section{Introduction}

Since stars are observed to form in molecular clouds (MCs), the properties of these stellar nurseries are expected to be closely tied to the processes of star formation.  It has been a continuing goal of astrophysicists to thoroughly characterize the structure of MCs and to understand how this structure relates to protostellar collapse.  The realization since at least the 1980s that MCs of similar masses can have drastically different rates of star formation gives further impetus to fulfilling this objective.  

The \mad~Molecular Cloud and the Rosette Molecular Cloud, as emblems of extreme contrasts in star formation activity, are good astrophysical laboratories for testing issues related to MC and star formation.  Also known as Maddalena's Cloud or the Maddalena-Thaddeus Cloud, \mad~is noteworthy for its low temperature ($\sim$ 10 K), high mass ($\sim 10^{5} \msun$), but low levels of star formation activity, compared to other MCs (Maddalena \& Thaddeus 1985).  In a series of papers in the 1990s, Lee, Snell \& Dickman (1991; 1994; 1996) mapped \mad~in \co{12} and \co{13} and studied its infrared properties.  Lee et al. confirmed the earlier result of Maddalena \& Thaddeus (1985) that the molecular cloud does not host massive star formation.  Through near-infrared imaging of point sources detected by the \emph{Infrared Astronomy Satellite} (\emph{IRAS}), they suggested that low-mass star formation is occurring primarily toward the boundaries of the cloud, and deduced that it is a remnant cloud that achieved its properties from past episodes of massive star formation.  Only recently were observations presented that indicate low-mass star formation is occurring in the main body of the cloud (Megeath et al. 2009).  The Rosette Molecular Cloud stands in stark contrast to \mad.  The well-studied MC contains the OB association NGC 2244 whose stars excite the Rosette Nebula which, in turn, may be triggering star formation elsewhere in the cloud (Heyer et al. 2006).  Mid-infrared, near-infrared, and X-ray observations have revealed the presence of several embedded clusters within the Rosette (e.g., Phelps \& Lada 1997; Rom\'{a}n-Z\'{u}\~{n}iga et al. 2008; Ybarra et al. 2013).  

Thus, the Rosette and \mad~represent two limits of star formation activity: on the one hand, the star formation rate (SFR) per unit mass in the Rosette is higher than the Galactic average (Williams \& McKee 1997), while \mad~does not contain even one O star, as is expected for most Galactic MCs of similar mass.

Does \mad~have such low levels of star formation because it is very young (a few Myr) and has only begun to form stars, as suggested by Maddalena \& Thaddeus (1985)?  Or is the MC in between episodes of massive star formation, as Lee et al. (1994) proposed?  The main disadvantage of the latter proposal is the lack of stars detected in \mad.  A study of the atomic hydrogen (\HI) gas associated with the cloud was more ambiguous in its conclusions about the cloud's evolutionary status (Williams \& Maddalena 1996).  Moreover, the arguments that \mad~is a relic of past star formation were made prior to the \emph{Spitzer} observations by Megeath et al. (2009), which show there is an embedded population of young stellar objects (YSOs) in the main body of the cloud.

In this and a following paper, I take a fresh look at \mad~to address the questions of the MC's evolutionary status and why it has such a low star formation activity.  In light of new observations since the 1990s, especially those of the \emph{Herschel} Space Observatory, as well as advances in numerical simulations of MC formation and evolution, a reexamination of this unusual molecular cloud is timely.  This paper addresses the \mad~mystery by comparing properties of its molecular gas---in particular, the column density distribution---with that of the well-studied Rosette Molecular Cloud, which is considered to be representative of MCs harboring high-mass star formation.  In \S 2, I review previous results that are germane to this study.  In \S 3 and \S 4, I discuss the data and methods used in this study. The results, including column density maps and CO-to-\htwo~conversion maps, are presented in \S 5.  The key findings are further discussed in \S 6 and summarized \S 7.

\section{Background}\label{sec:background}
In the 1990s, Lee et al. (1991; 1994; 1996) reported on a rigorous set of investigations into the molecular gas, star formation, and dust properties in \mad.  Using the FCRAO 14 m telescope, Lee et al. (1991) mapped \mad~in the \co{12} and \co{13} $J=1-0$ transitions.   They showed that with \co{12} antenna temperatures of less than 4 K across the cloud, \mad~is colder than most observed MCs. In the second paper, Lee et al. (1994) described how the large \co{12} line widths, $\sim 8~\kms$, might have resulted from prior star formation.  Their data indicate a complex velocity structure (e.g., shells and rings) that could be the vestiges of past massive star formation activity.   They estimated a mass of $1.1 \times 10^5 \msun$ from \co{13} observations and an assumption of local thermodynamic equilibrium (LTE).  To infer the total hydrogen column density, they derived a relation between \co{13} and \av, which they measured from star counts.  Adopting a distance to \mad~of 2.2 kpc and a diameter of 100.8 pc, Lee et al. (1994) determined a virial mass of $6.3 \times 10^5 \msun$.  They suggested that the factor of 6 discrepancy between the two mass estimates could be explained if the cloud is not in virial equilibrium but, rather, is undergoing expansion.  Nevertheless, even by the LTE mass estimate, \mad~is more massive than many other nearby Galactic, star-forming MCs (e.g., Imara \& Blitz 2011).  

\begin{figure*}
\includegraphics[width=\textwidth]{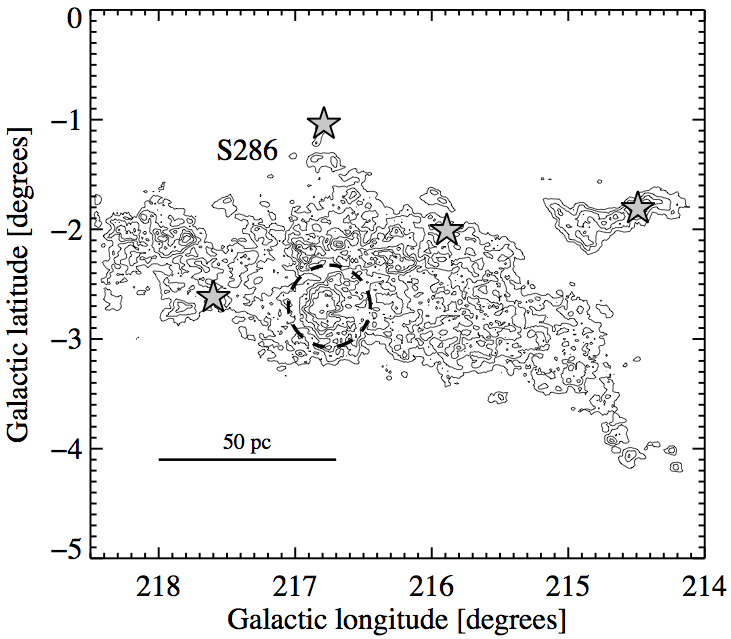}
\caption{Integrated \co{12} emission of the \mad~molecular cloud.  Contour levels are from 2 to 12 \counits~and spaced 2 \counits~apart.  The gray stars indicate the locations of \emph{IRAS} point sources identified by Lee et al. (1996).  The dashed circle shows the region where Megeath et al. (2009) identified 71 protostars and YSOs.  S286 is a distant \HII~region unrelated to the molecular cloud. \label{fig1}}
\end{figure*}

Lee et al. (1996) utilized infrared observations to infer the dust and low-mass star formation properties of \mad.  They derived a correlation between dust and gas emission reminiscent of that observed in nearby dark clouds.  They measured a far-infrared (FIR) luminosity to molecular gas mass ratio of $L_{\t{FIR}}/M_{\t{H}_2}<0.7 L_\odot/M_\odot$.  This ratio, which is often used as a proxy for star formation efficiency, is close to $\sim 1$ for MCs in the solar neighborhood (e.g., Scoville \& Good 1989) and is $\sim 10$ for the Rosette (Heyer et al. 2006).  Lee et al. (1996) examined objects in the \emph{IRAS} Point Source Catalog in the vicinity of the cloud, and with follow-up near-infrared observations in the $J$, $H$, and $K$ bands, they identified four of the \emph{IRAS} sources as possible sites of low-mass star formation (Figure \ref{fig1}).  Altogether, they estimated that 14 young stellar objects (YSOs) are located at these sites.  Notably, none of these sources are located in the main body of the cloud.  Two are associated with satellite clouds to the north of the main MC, and the other two are located near the boundary of CO emission.  In this study, Lee et al. (1996) also pointed out that infrared emission in the north-east region of \mad~is contaminated by emission from the more distant \HII~region S286 (Figure \ref{fig1}). 

A little more than a decade passed before Megeath et al. (2009) found evidence of low-mass star formation within the main body of \mad. This group took observations with the \emph{Spitzer} Telescope's Infrared Array Camera (IRAC) and Multiband Imaging Photometer for \emph{Spitzer}.  Using color-color selection criteria, they identified 74 YSOs corresponding to a peak in \co{} emission located near $(l,b)=(216.7^\circ,-2.75^\circ)$, (see Figure \ref{fig1}).  They noted that the surface density distribution of YSOs in \mad~is similar to the distribution in Taurus and much less than the surface density of YSOs in Orion A.  Curiously, \mad~is roughly 10 times as massive as Taurus.

The remainder of this work will explore the molecular gas distribution of \mad, as revealed in rich detail by \emph{Herschel}.  I will derive and make detailed comparisons between the \htwo~column density maps of the \mad~and Rosette MCs, in order to probe the evolutionary state of the former.  I will also address the question of whether the CO-to-\htwo~conversion factor, the so-called $X$-factor, differs depending on local cloud conditions.  While \htwo~is the most abundant molecule in the interstellar medium, it cannot be observed directly at the cold temperatures of molecular clouds because it does not have a permanent dipole moment.  Therefore, alternative tracers of molecular gas must be used to infer the properties of molecular clouds.  The second most abundant molecule, CO, is commonly used, especially in external galaxies.  To convert integrated CO emission, \ico, into \htwo~column density, $N(\htwo)$, one can use the $X$-factor,
\begin{equation}\label{eq:xco}
\xco=\frac{N(\htwo)}{\ico}~\xunits.
\end{equation}
Bolatto et al. (2013) discussed in detail the independent techniques employed to determine \xco.  One involves using tracers of column density, such as dust or optically thin molecular lines.  Historically, this was the first method used (Dickman 1975) and is justified since dust is observed to be well-integrated with gas (e.g., Boulanger et al. 1996) and since the \htwo~molecule forms most efficiently on dust grains. Dickman (1975) observed the \co{13} emission from 68 Galactic clouds and compared this to optical extinction measurements.  If a typically observed ratio of $I(\co{13})$ to $I(\co{12})$ is applied to those observations, this yields $\xco=2.2\times 10^{20}$.  The method used to derive \xco~in this analysis belongs to the same category as that initiated by Dickman (1975).  In this case, however, dust far-infrared \emph{emission} (as opposed to dust \emph{extinction}) is used to trace the total hydrogen distribution.  To convert from dust to total gas to \htwo~requires certain assumptions, including a gas-to-dust ratio (see \S\ref{sec:methods}).

Another type of \xco~determination requires the potentially problematic assumption that MCs are in virial equilibrium, and the third compares CO emission with $\gamma$-ray emission produced by cosmic-ray interaction with \htwo~molecules.  While there is general accord among the different techniques to measure \xco, variations in \xco~have nevertheless been observed.  Bolatto et al. (2013) propose a Galactic conversion factor of $2\times 10^{20}$  with $\pm$ 30\% uncertainty, noting that departures from this value should be anticipated.    Dame et al. (2001) showed that average value of \xco~in the Solar neighborhood is $1.8\pm 0.3 \times 10^{20}~\xunits$ and varies as a function of Galactic latitude.  The conversion factor has been found to vary in spatially resolved Galactic MCs (e.g., Pineda et al. 2008, 2010) and also in dwarf galaxies (e.g., Imara \& Blitz 2007; Leroy et al. 2007) that have low metallicity with respect to the Galaxy.  It is one of the aims of this study to address the unsettled questions of how and what other environmental factors drive changes in \xco.

\section{Data}
To determine the quantity and distribution of molecular gas across the molecular clouds, I use far-infrared data from the \emph{Herschel} Science Archives.  Images in the PACS wavelengths (70 \micron~and 160 \micron) and SPIRE wavelengths (250 \micron, 350 \micron, and 500 \micron) that had passed Level 2 processing were available for regions of both \mad~(OT2 PI: J. Kaufmann) and the Rosette (Motte et al. 2010).  For \mad, only the SPIRE images are used for this analysis, since they cover the entire region of interest.  (Given the weakness of the PACS emission at the low temperatures characteristic of \mad, its exclusion is not expected to significantly modify the results of the fits.  As will be seen in \S \ref{sec:results}, the global properties derived for \mad~are consistent with those arrived at in previous studies using independent methods.)  For the Rosette, both the PACS and SPIRE data are used, though they only cover a partial region (roughly $1.5^\circ \times 1.5^\circ$) of the cloud to the east of the OB cluster NGC 2244.  

Both clouds lie roughly two degrees below the Galactic midplane.  \mad~is centered at Galactic coordinates  $(l,b)=(216.5^\circ,-2.5^\circ)$; a distance of 2.2 kpc is adopted (Lee et al. 1991).  The Rosette is located at $(l,b)=(207^\circ,-2^\circ)$, at a distance of 1.33 kpc (Lombardi et al. 2011). 

The CO data were originally obtained with the FCRAO 14 m telescope and were graciously provided by Y. Lee (originally published in Lee et al. 1994) for \mad, and by M. Heyer for the Rosette (see Heyer et al. 2006).  For \mad, the spatial and velocity resolutions of the \co{12} data cube are 50\arc~and 0.65 \kms.  For the Rosette, the spatial and velocity resolutions are 20\arc~and 0.127 \kms.  The average rms noise levels in the \co{12} data cubes are 0.3 and 1.0 K, respectively.

\section{Methods}\label{sec:methods}
In this section, I describe the procedure used to determine the dust temperature, $T$, and column density distributions.  

\textbf{Modified blackbody fits} --- 
The starting assumption is that the far-infrared emission in the MCs arises from cold dust which may be treated as a modified blackbody of the form:
\begin{equation}
I_\nu = B_\nu(T)[1 - \e^{-\tau_\nu}]
\end{equation}
where $B_\nu$ is the Planck function, given by:
\begin{equation}
B_\nu = \frac{2h\nu^3}{c^2}\frac{1}{\e^{h\nu/kT} - 1}
\end{equation}
The dust optical depth, $\tau_\nu$, is given by
\begin{equation}\label{eq:tau}
\tau_\nu= \mu m_\H \kappa_\nu N(\H)\times(1/100) 
\end{equation}
where $\mu = 1.36$ is the helium correction, $m_\H$ is the mass of a hydrogen atom, $N(\H)$ is the total hydrogen gas column density, and $\kappa_\nu$ is the dust opacity in units of $\t{cm}^2$ $\t{g}^{-1}$.  The factor of $1/100$ in Equation \ref{eq:tau} is the assumed dust-to-gas ratio.  The dust opacity depends on the properties and distribution of the dust and is treated as a power law (following Beckwith et al. 1990),
\begin{equation}
\kappa_\nu = 0.1 \left(\frac{\nu}{1000~\mathrm{GHz}}\right)^\beta,
\end{equation}
where $\beta\sim 2$ is the dust emissivity index.

Each pixel in a PACS or SPIRE map represents the flux at a certain wavelength, with three unknown values of interest: $T$, $N(\H)$, and $\kappa_\nu$ (or $\beta$).   The first step is to convolve all the images to a common resolution, that is, to the poorest resolution of 36\arc, corresponding to the SPIRE 500 \micron~data.  Next, each image is projected onto a grid having the same number of elements, pixel size, and units.  (The PACS and SPIRE data are originally in units of Jy/pixel and MJy/steradian, respectively.)   To convert the observed flux of the images into intensity, it is necessary to take into account the beam size at different wavelengths.  Finally, the data are fit on a pixel-by-pixel basis using the MPFITFUN procedure from the Markwardt IDL Library (Markwardt 2009), leaving  $T$ and $N(\H)$ as free parameters and fixing the dust emissivity index at $\beta = 2$.  The temperature was limited to the range 3 to 55 K; $N(\H)$ was restricted to vary between $0$ and $10^{24}$ \cm.

\textbf{The X-factor} ---
For each molecular cloud, the \xco~map is determined from the ratio of the $N(\htwo)$ and integrated \co{12} (hereafter, CO) intensity maps (Equation \ref{eq:xco}).  The \htwo~column density is determined by subtracting the \HI~contribution from the total hydrogen column density, using $N(\H)=N(\HI)+2N(\htwo)$.  The average \HI~column density, as measured from the LAB \HI~survey, is $\sim 1.0\times 10^{21}~\cm$ toward \mad~and $\sim 1.7\times 10^{21}~\cm$ toward the Rosette.  The integrated \co{12} intensity, \ico, is derived by integrating the CO emission over a certain velocity range.   The boundaries of the clouds are defined by the 3-$\sigma_\t{RMS}$ level.  For \mad, the CO emission is integrated from 17 to 37 \kms, and the 1-$\sigma_\t{RMS}$ level is 0.5 K \kms.  For Rosette, the velocity range is 5 to 25 \kms; the 1-$\sigma_\t{RMS}$ level is 0.6 K \kms.

Since the CO and \emph{Herschel} data sets have different resolutions, they are regridded onto maps having the same pixel size before the ratio is taken.  To achieve similar linear resolutions at the adopted distances to the clouds, thereby facilitating comparison between them, the $N(\htwo)$ maps derived from \emph{Herschel} data are smoothed to the resolution of the CO data 50\arc~($=0.53$ pc at a distance of 2.2 kpc); and the Rosette maps are smoothed to an angular resolution of 80\arc~($=0.52$ pc at 1.33 kpc).

\section{Results}\label{sec:results}
\subsection{Distribution of molecular gas}\label{sec:distribution}
\begin{figure*}
\includegraphics[width=\textwidth]{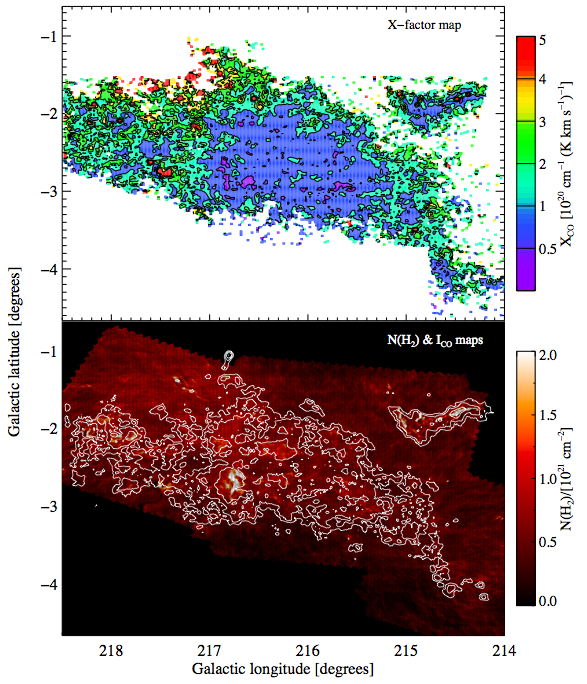}
\caption{\textbf{\mad}: Bottom panel -- $N(\htwo)$ map overlaid with \co{12} contours at 2, 4, 8, and 16 \counits.  The resolution of the maps are 36\arc~and 50\arc, respectively.  Top panel -- \xco~map. To derive \xco, the  $N(\htwo)$ map is convolved to the same resolution as the \co{12} map. \label{fig2}}
\end{figure*}

\begin{figure*}
\includegraphics[width=\textwidth]{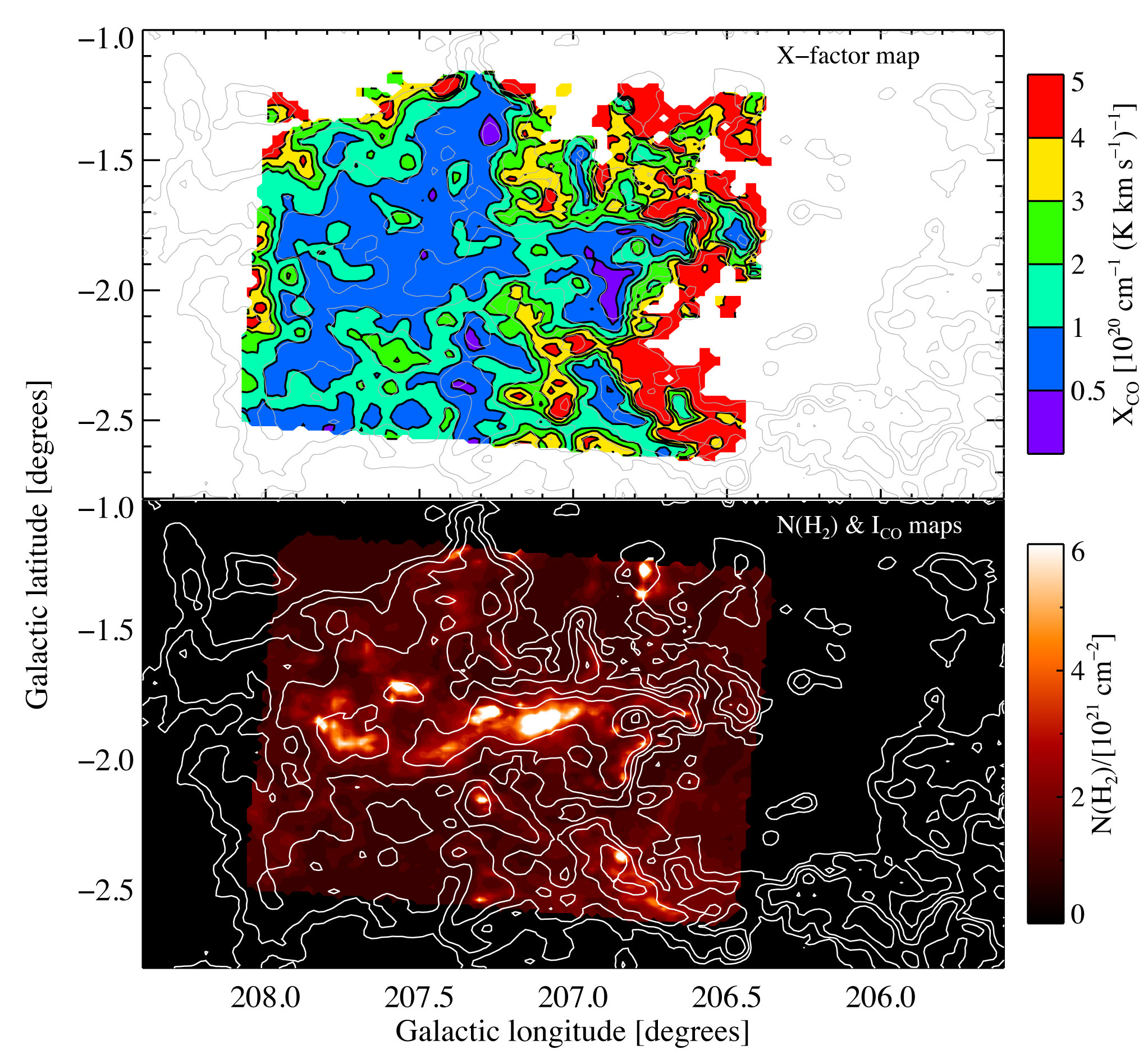}
\caption{\textbf{Rosette}: Bottom panel: $N(\htwo)$ map overlaid with \co{12} contours at 2, 4, 8, and 16 \counits.  The resolution of the maps are 36\arc~and 80\arc, respectively.  Top panel -- \xco~map with CO contours overlaid in gray.     \label{fig3}}
\end{figure*}

Figures  \ref{fig2} and \ref{fig3} show the \htwo~column density maps derived from \emph{Herschel} data, with CO contours overlaid.  The average column density of \htwo~gas in \mad~is $\sim 7.8 \times 10^{20}$ \cm, and the total molecular mass is $1.2 \times 10^5~\msun$, very close to the LTE mass calculated by Lee et al. (1994) using \co{13} measurements.  The bottom panels of Figures \ref{fig2} and \ref{fig3} show that \ico~closely follows the distribution of \htwo~gas in both molecular clouds.  While the peaks in CO emission generally correlate with peaks in $N(\htwo)$, the $N(\htwo)$ maps reveal much more small-scale structure.  Moreover, in the \mad~molecular cloud, there are a number of small \htwo~density enhancements with little associated CO emission.

For instance, there is a knot located at $l=215.6^{\circ}$, $b=-3.4^{\circ}$ that has a peak column density of $\sim 7\times 10^{21}$ \cm, which is above the star formation ``threshold'' that has been posited by some authors (e.g., Lada et al. 2010; see \S \ref{sec:discussion}).  However, the average density in this knot is $\sim 1.6\times 10^{21}$ \cm.  At a size of roughly 0.8 pc, this corresponds to a particle volume density of $n\approx 500~cm^{-3}$.  There are similar high-density knots that have peak column densities above the star formation ``threshold'' observed in other molecular clouds, with $\ico\sim 2$ \counits~located at  $l=215.5^{\circ}$, $b=-3.7^{\circ}$ and  $l=217.6^{\circ}$, $b=-2.6^{\circ}$.  By comparison, the region around  $l=216.7^\circ$, $b=-2.75^\circ$, where Megeath et al. (2009) identified YSOs, peaks at a column density of $\sim 2\times 10^{21}$ \cm~and $\ico\sim 12$ \counits. In the satellite cloud located to the north-west of the main MC, there are knotty, filamentary-like structures that are all bounded by CO contours between $\sim$ 2 and 4 \counits.  The nearly horizontal filament to the far west of the satellite, the location of \emph{IRAS} point sources (Lee et al. 1996), is roughly 13 pc $\times$ 1.5 pc.  It has a column density of $\sim 5\times 10^{21}~\cm$ and $\ico\sim 10$ \counits.  Extending along the eastern border of the satellite is a $\sim 12$ pc $\times$ 1.5 pc structure with $N(\htwo)\sim 2\times 10^{21}~\cm$ and $\ico\sim 6$ \counits. In \S \ref{sec:regional}, these high-density regions, both star-forming and non-star-forming, will be examined more closely. 

Within the mapped region of the Rosette (Figure \ref{fig3}), $\avg{N(\htwo)}\approx 1.8 \times 10^{21}~\cm$, and the total molecular mass is $3.7 \times 10^4~\msun$.  Considering that all of the molecular gas has not been counted, the value estimated here is in good agreement with previous determinations (e.g., Heyer et al. 2006; Imara \& Blitz 2011).  Heyer et al. (2006) found a mass nearly 4 times higher, but they assumed a greater distance to the Rosette (1600 pc) than the distance used here and so effectively integrated over a larger area.  Using this distance would increase the mass calculated here only by a factor of $(1600/1330)^2=1.4$.  Figure \ref{fig3} indicates that for the area mapped, there are fewer places (compared to \mad), where there are peaks in $N(\htwo)$ without corresponding peaks in \ico.  

\begin{figure}
\plotone{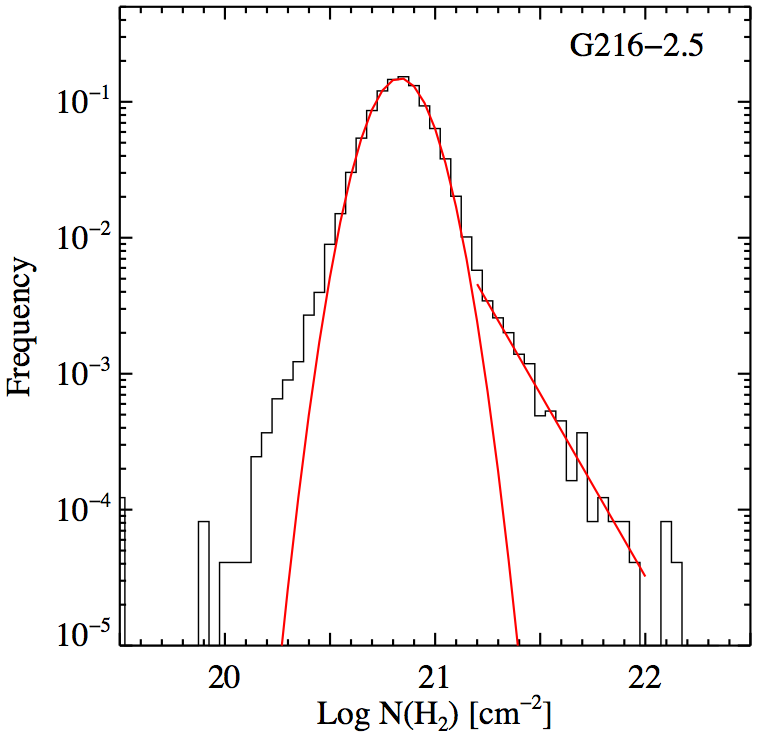}
\caption{PDF of \htwo~column densities in \mad.  Only pixels corresponding to regions where $\ico \ge 3-\sigma_\t{RMS}$ are used.  The PDF peaks at $N(\htwo)=6.8\times 10^{20}~\cm$ and has a power-law tail whose slope in log-log space is $-6.2$.  These and other parameters are summarized in Table 1. \label{fig4}}
\end{figure}

\begin{figure}
\plotone{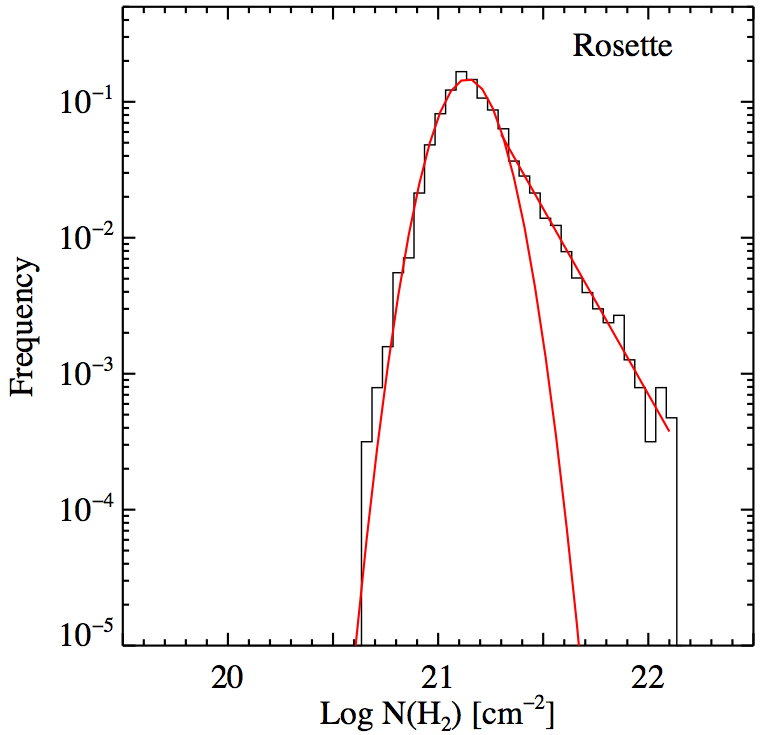}
\caption{PDF of \htwo~column densities in the Rosette.  Only pixels corresponding to regions where $\ico \ge 3-\sigma_\t{RMS}$ are used.  The PDF peaks at $N(\htwo)=1.4\times 10^{21}~\cm$ and has a power-law tail whose slope in log-log space is $-6.3$.  These and other parameters are summarized in Table 1.\label{fig5}}
\end{figure}

To get another perspective on how high-density gas is distributed in the MCs, I consider the probability distribution function (PDF) of the column density and the cumulative mass fraction (CMF) distribution. Figures \ref{fig4} and \ref{fig5} display the normalized PDFs of the MCs in log-log space, in which a log-normal distribution would appear as a parabola and a power-law as a straight line.   Indeed, the PDFs take on the classical log-normal shape with a power-law tail at high column densities observed in other MCs (see e.g., Kainulainen et al. 2009; Froebrich \& Rowles 2010; Lombardi et a1. 2014) and predicted in theoretical studies (e.g., Federrath \& Klessen 2013; Clark \& Glover 2014; Ward et al. 2014).  For a range of $N(\htwo)$ values before the power-law tail begins, the logarithm of data points are fit by a Gaussian
\begin{equation}
f(\log~N(\htwo);\mu,\sigma)=\frac{1}{\sigma\sqrt{2\pi}}~e^{-\frac{(\log~N(\t{H}_2)  - \mu)^2}{2\sigma^2} },
\end{equation}
where $\mu$ is where the distribution peaks, effectively providing the average $N(\htwo)$ in the MC, and $\sigma$ is the dispersion.  The high-extinction portions of the PDFs are fit by a linear function (in log-log space).  The Gaussian fit is robust: changing the range of data points fitted does not significantly alter the results, which are presented in Table 1.  The \mad~PDF peaks at $N(\htwo)=6.8\times 10^{20}~\cm$  and has an excess over log-normal at low column densities.  The PDF of the Rosette, which peaks at $N(\htwo)=1.4 \times 10^{21}$ \cm, is truncated at low column densities the low-density gas in the cloud is not fully represented in the \emph{Herschel} map.  Also note that the \mad~PDF has a slightly shallower power-law tail than that of the Rosette.  Possible implications of these results will be discussed in \S \ref{sec:discussion}.

\begin{table}\centering
\begin{center}
\begin{tabular}{lcc}
\multicolumn{3}{c}{Table 1: Molecular cloud PDF parameters.}\\
\tableline\tableline
Parameter                       &    \mad        & Rosette        \\
\tableline
$(\t{Log}~N(\htwo))_{\t{peak}}$  & $20.8\pm 0.1$            & $21.1\pm 0.1$    \\
$N(\htwo)_{\t{peak}}$            & $6.8\times 10^{20}~\cm$  & $1.4\times 10^{21}~\cm$   \\
$m$                             & $-6.2\pm0.1$   & $-6.3\pm 0.1$  \\
TP                              & $2.0\times 10^{21}~\cm$        & $2.0\times 10^{21}~\cm$        \\
\tableline
\end{tabular}
\begin{tablenotes}
\footnotesize
\item
\textbf{Note.} $(\t{Log}~\av)_{\t{peak}}$ and  $\av_{\t{peak}}$ are the values where the PDF peaks.  $m$ is the slope of the high-\av~power-law tail of the PDF. The transition point (TP) is the value of $N(\htwo)$ where the PDF transitions from log-normal to power-law.
\end{tablenotes}
\end{center}
\end{table}

\begin{figure}
\plotone{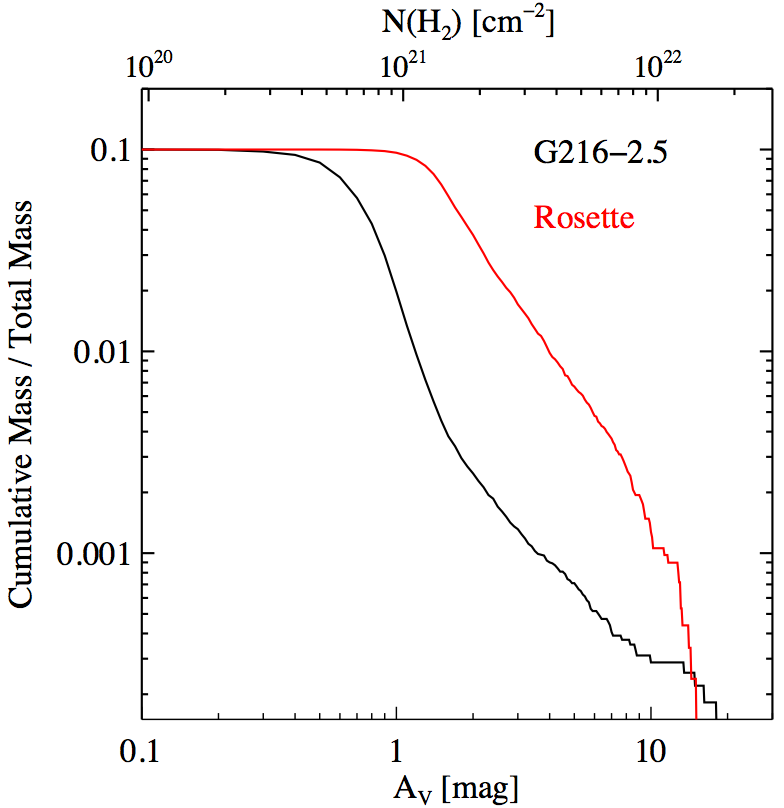}
\caption{The normalized cumulative mass profiles as a function of $N(\htwo)$ (equivalently, visual extinction \av~due to \htwo) of the \mad~and Rosette MCs.\label{fig6}}
\end{figure}

The plot of normalized CMFs displayed in Figure \ref{fig6} has horizontal axes in units of both column density and visual extinction, using the conversion $N(\htwo)/ \av = 0.94\times 10^{21}$ mag$^{-1}$ cm$^{-2}$.  The CMFs show that over a wide range of \av, spanning $\sim 0.5$ to 15 mag, the Rosette has a higher fraction of its mass locked up in dense gas than does \mad.  It is noteworthy that the \mad~CMF falls rapidly at low extinctions and then becomes shallower starting around 1 -- 2 mag, revealing yet again that most of its mass is in the form of low-density gas.  The total molecular mass of \mad~and the mapped region of the Rosette are $1.2\times 10^5~\msun$ and $3.7\times 10^4~\msun$, respectively.

There may be appreciable region-to-region variations in \mad~that impact its ability to form stars or that are the result of past episodes of star formation.  Thus, in \S \ref{sec:regional}, I will divide the MC into different regions and examine their individual PDFs.  Consequences of using different values of $\beta$ are discussed in \ref{sec:beta}.

\subsection{The $X$-factor}\label{sec:xfactor}
The top panels of Figures \ref{fig2} and \ref{fig3} show the spatial variations of \xco~across \mad~and the Rosette.   In both clouds, \xco~is determined from Equation 1 only for regions where \ico~is greater than the 3-$\sigma_\t{RMS}$ level.  In \mad, \xco~increases from $\sim 0.5\times 10^{20}$ in the central region to $\sim 5\times 10^{20}$ at the periphery.  In the region of \mad~where Megeath et al. (2009) identified embedded sources, \xco~is slightly smaller, $\sim 1.4 \times 10^{20}$, than the typical Galactic value recommended by Bolatto et al. (2013).  Figure \ref{fig3} shows that for the Rosette, too, \xco~ranges from $\sim 0.5$ to $\sim 5\times 10^{20}$.  Using the coordinates of young clusters from the Ybarra et al. (2013) study, YSOs in the Rosette tend to be located in regions that have, on average, higher values of \xco~($\sim 2.6 \times 10^{20}$) than \mad.   

\begin{figure}
\plotone{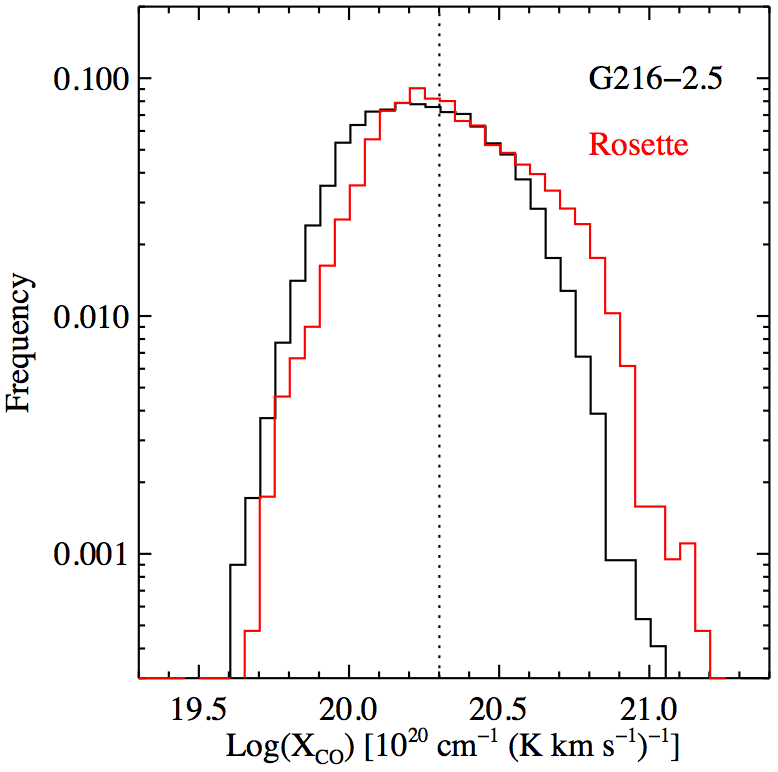}
\caption{Histograms of Log(\xco).  The mean values for \mad~and the Rosette, respectively, are $2.2\times 10^{20}$ and $2.8\times 10^{20}$.  The dotted line indicates the typical Galactic value of $2\times 10^{20}$.  \label{fig7}}
\end{figure}

The histograms in Figure \ref{fig7} show that the two molecular clouds have similar distributions of \xco.  For \mad, the mean value is $\avg{\xco}=(2.2\pm 1.3)\times 10^{20}$; for the Rosette, $\avg{\xco}=(2.8\pm 2.1)\times 10^{20}$; (see Table 2).  Uncertainties in measurements of \xco, $\sigma_{X_{\t{CO}}}$, are determined by propagating the uncertainties in the $N(\htwo)$ and CO maps.  

Another way to determine the global value of \xco~is to take the ratio of the mean cloud properties: i.e., $\langle N\htwo \rangle/\langle \ico \rangle$.  This method is perhaps more suitable for galaxy-wide studies of MCs, in which the goal is often to determine the masses of whole clouds.  For \mad, $\langle\xco\rangle = \langle N\htwo \rangle/\langle \ico \rangle = (1.8\pm 1.5)\times 10^{20}$. For the Rosette, $\langle N\htwo \rangle/\langle \ico \rangle = (2.5\pm 2.4)\times 10^{20}$.  These mean values are somewhat lower than those cited in the above paragraph, since the former are more influenced by large values of $\xco$ at the edges of the clouds.  Nevertheless, within the stated uncertainties, the global values of \xco~estimated here are in agreement with one another.

\begin{table}\centering
\begin{center}
\begin{tabular}{ccc}
\multicolumn{3}{c}{Table 2: Derived \xco~$[10^{20}~\xunits]$.}\\
\tableline\tableline
\xco    & \mad    & Rosette \\
\tableline
Mean           &  $2.2 \pm 1.3$            &   $2.8\pm 2.1 $  \\
\ico~vs. \av   &  $0.46 + \frac{5.6}{\ico}$ & $0.87 + \frac{9.6}{\ico}$         \\
\tableline
\end{tabular}
\end{center}
\end{table}

\begin{figure}
\plotone{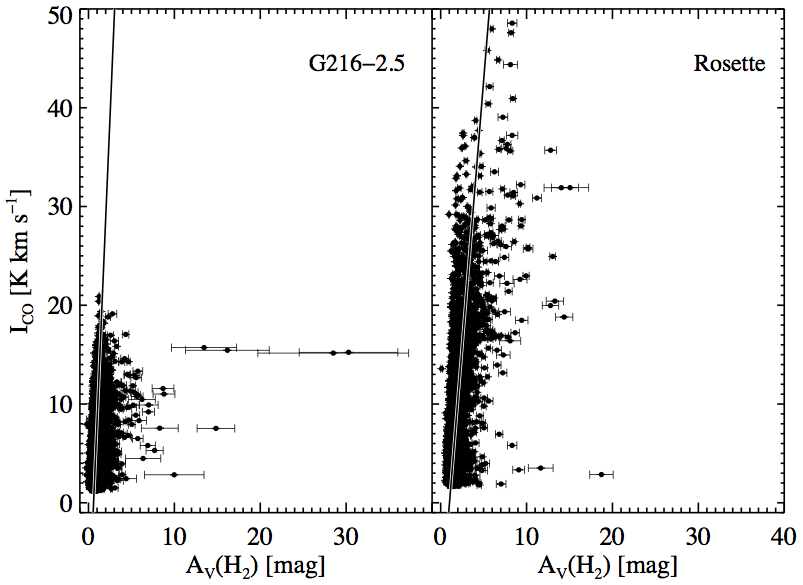}
\caption{\ico~versus \av: Each data point represents values of a pixel in Figures \ref{fig2} and \ref{fig3}.  Overplotted is the best fit from Bayesian linear regression: $\av = m\ico +b+\epsilon$, where $m$ is the slope, $b$ is the \av~below which no CO is expected to be detected, and $\epsilon$ is the intrinsic scatter about the regression.  For \mad, $\av = 0.049\ico + 0.59$ and $\epsilon=0.06$; (equivalently, $\xco=[0.46+5.6/\ico]\times 10^{20}$).  For the Rosette, $\av = 0.093\ico + 1.0$ and $\epsilon= 0.46$; ($\xco=[0.87 + 9.6/\ico]\times 10^{20}$).     \label{fig8}}
\end{figure}

A third way to investigate \xco~is to compare \ico~and \av~(e.g., Pineda et al. 2008; Lee et al. 2014).  In Figure \ref{fig8}, \ico~is plotted as a function of \av, the extinction due to \htwo.  As noted in \S \ref{sec:methods}, the column density images of the two MCs have been regridded to maps having the same resolution as the \ico~maps.  The error bars in Figure \ref{fig8} are from the dispersion in each pixel of the $N(\htwo)$ maps.  Error bars for \ico~are not shown, but the 1-$\sigma$ uncertainty in the CO emission is taken into account in the following.  A Bayesian linear regression analysis is performed on the data, according to
\begin{equation}\label{eq:linear}
\av=m\ico + b + \epsilon,
\end{equation}
where the slope, $m$, is related to $\xco$, $b$ is the value of \av~below which no CO emission is expected to be detected, and $\epsilon$ is the intrinsic scatter around the regression.  The fitted parameters are listed in Table 2.  For \mad, $m=(0.049\pm 0.001)$ mag $(\counits)^{-1}$ and $b=(0.593\pm 0.003)$ mag. For the Rosette,  $m=(0.093 \pm 0.001)$ mag $(\counits)^{-1}$ and $b=(1.03\pm 0.01)$ mag.  If the fit to the Rosette cloud were made over the same range of \ico~values as for \mad, (i.e., for $\ico <20$ \counits), the resulting fit would be slightly shallower.  Varying the range of \av~values fitted, for either cloud, does not significantly change the slope of the fit.  It is noteworthy that for \mad, $b$ is close to 0.5, the lower limit for detection in dust extinction observations (e.g., Kainulainen et al. 2009), and is also near the column density at which atomic hydrogen begins to self shield from UV radiation and form \htwo.  The best fit for $b$ in the Rosette is somewhat high because low-column density regions of the cloud were not used in this analysis, and it most likely does not represent a true detection threshold. In their study of \xco~in the Perseus MC, Lee et al. (2014) say that $m\times 0.94\times 10^{21}$ is essentially \xco.  Here, I do not exclude the functional dependence of \xco~on \ico.  Thus, for \mad, the best fit parameters yield $\xco = (0.46 + 5.6/\ico)\times 10^{20}$.  Using the average CO integrated intensity across the cloud, $\avg{\ico}=4.4$ \counits, gives $\xco\approx 1.7\times 10^{20}$.  For the Rosette, the best fit parameters and  $\avg{\ico}=7.2$ \counits, yield $\xco = (0.87 + 9.6/\ico)\times 10^{20}\approx 2.2\times 10^{20}$.

\begin{figure*}
\includegraphics[width=\textwidth]{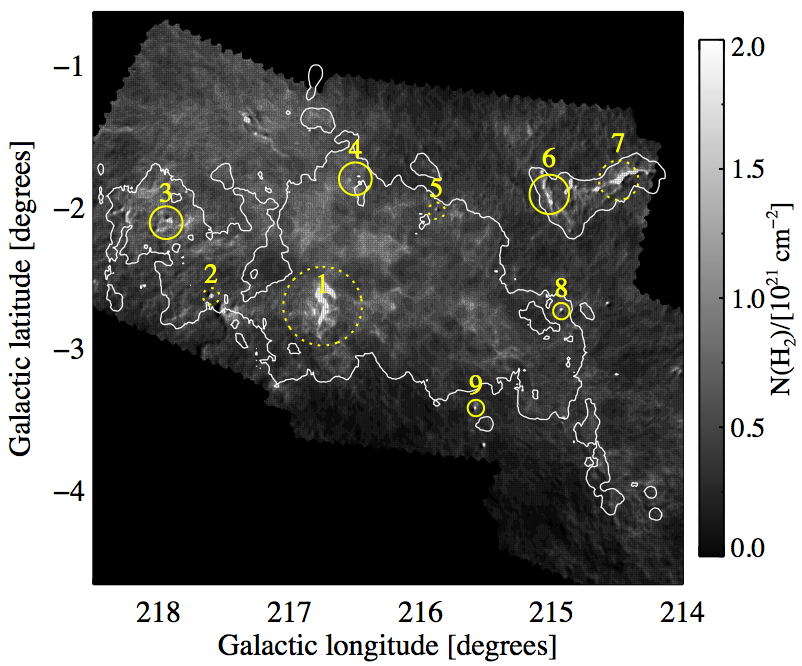}
\caption{High-column-density regions selected for closer examination.  The resolution of the $N(\htwo)$ map is 36\arc, the same as Figure \ref{fig2}.  The dashed circles circumscribe regions where star formation has been identified.  Solid circles show high-$N(\htwo)$ regions where star formation has not been detected.  All of the pixels within each circle are included in the corresponding plots shown in Figures \ref{fig10} and \ref{fig11}.\label{fig9}}
\end{figure*}

\subsection{Regional variations in \mad}\label{sec:regional}
Much of the scatter in the $X$-factor and the \ico-\av~relation is due to region-to-region variations in the physical environments across a given cloud.  Are some regions more conducive to star formation than others?  Why?  Examining these parameters in individual regions of \mad~could provide insight into the cloud's dearth of star formation relative to other MCs.  In particular, one might ask the question: do regions where there is evidence of low-mass star formation activity exhibit behavior similar to an actively, star-forming MC, such as the Rosette?  To begin addressing this question, I select nine regions of \mad, with the help of several plots and maps, for closer study.  Four of the regions (1, 2, 5, and 7) are selected due to evidence of low-mass star formation within them (see \S \ref{sec:background}).  The other regions are selected for having high column densities similar to the star-forming sites.  Figure \ref{fig9} re-displays the column density map of \mad~with the regions circled and numbered.

\begin{figure*}
\includegraphics[width=\textwidth]{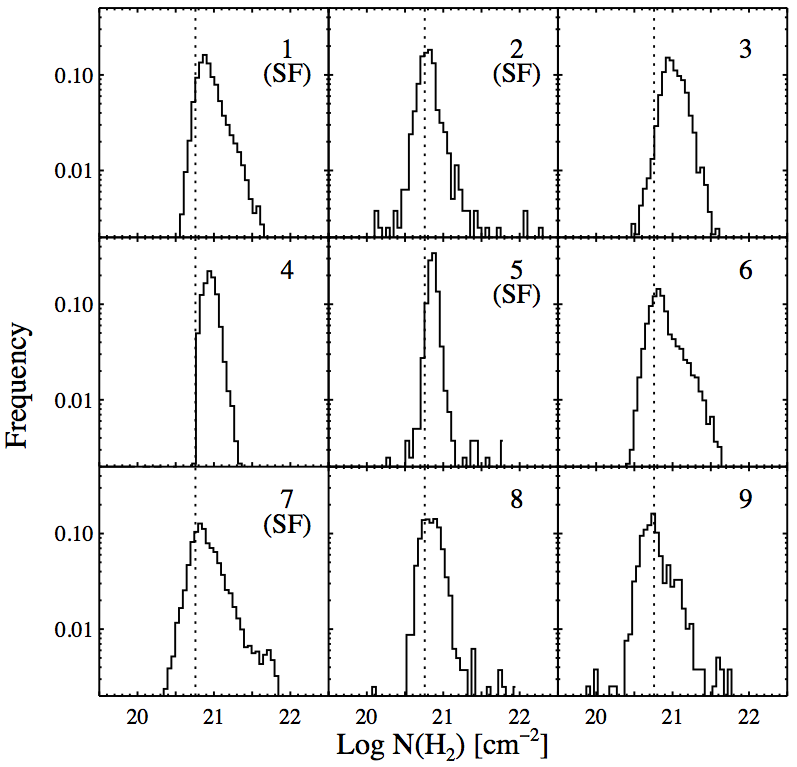}
\caption{PDFs of regions indicated in Figure \ref{fig9}. Regions where star formation activity has been identified are labeled (SF).  For reference, the dotted line indicates the location of the peak of the PDF of the entire cloud, $(\t{Log} N(\htwo))_{\t{peak}} = 20.8$. \label{fig10}}
\end{figure*}

The PDFs of the individual regions are shown on the same scale in Figure \ref{fig10}.  The histograms were generated from the high-resolution $N(\htwo)$ map prior to smoothing; each plot contains all the pixels within the corresponding circle in Figure \ref{fig9}.  The smallest regions have a radius of $225\arc$ (2.4 pc) and contain an average of 812 data points.  The size of the largest region, 1, was defined to include all of the embedded sources detected by Megeath et al. (2009); the radius is $1080\arc$ (11.5 pc), and it contains 18,679 pixels.  With the exception of region 9, each PDF peaks at a slightly higher column density than that of the PDF of the entire cloud ($\log~N(\htwo)=20.8$; Figure \ref{fig4}), but none of them peaks at $\log~N(\htwo)=21.1$, where the Rosette PDF peaks (Figure \ref{fig5}).   Plots 1, 2, 6, and 7 are perhaps the best representatives of regions displaying some degree of log-normality with power-law tails.  

\begin{figure*}
\includegraphics[width=\textwidth,height=7.04in]{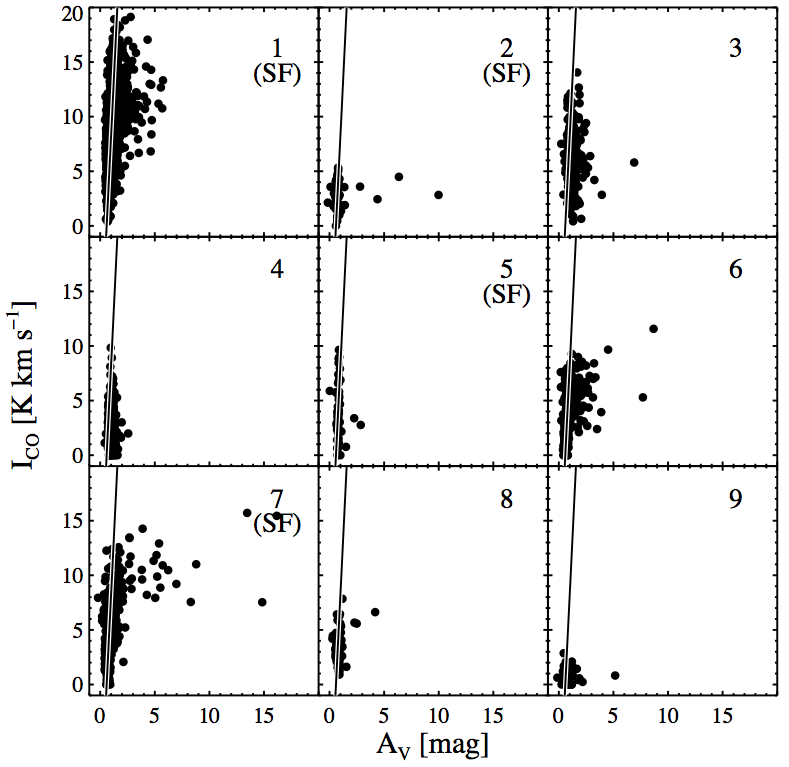}
\caption{\ico~vs. \av~plots of regions indicated in Figure \ref{fig9}. Regions where star formation activity has been identified are labeled (SF).  Overplotted for reference is the best fit to the data for the entire cloud (black line).   \label{fig11}}
\end{figure*}

The \ico-\av~plot for each individual region is shown with the linear fit derived for the whole MC, for comparison, in Figure \ref{fig11}.  These plots are created from the map in Figure \ref{fig2} and contain between 61 and 1440 data points.  Regions 1, 2, and 7, where there is evidence for star formation, show strong evidence for \co{12}~``saturating'' at high extinctions.  Interestingly regions 6 and 3 (less so) show similar saturation, although embedded sources have not yet been observed to be forming in the dense gas there (see \S \ref{sec:discussion} for more discussion).  Another noteworthy feature is the near-verticality of the plots corresponding to regions 4, 5, and 8.  In terms of Equation \ref{eq:linear}, this corresponds to a slope equal to zero; in other words, these non-star-forming regions appear to have a constant $X$-factor.

\subsection{Varying $\beta$ and uncertainties}\label{sec:beta}
The major source of uncertainty in this analysis arises from the modified blackbody fitting described in \S \ref{sec:methods}.  Single values of $N(\htwo)$, $T$, or the emissivity index $\beta$ along each line-of-sight do not reflect reality.  Apparent variations in temperature may actually be variations in $\beta$.  Using different values of $\beta$ in the the modified blackbody fitting would result in different outcomes of the $N(\htwo)$ maps and, thus, the measurements of \xco.  As $\beta$ decreases, dust temperature increases and $N(\htwo)$ decreases, resulting in a smaller overall $X$-factor and less massive clouds.  

In this analysis, after testing different values of $\beta$, I use a fixed value of $\beta = 2$ and found this to be a valid choice, since it yields average MC column densities and total masses in agreement with independent methods of measuring these quantities.  For instance, if I had adopted a value of $\beta=1.8$, the average temperature in each MC would be $\sim 1$ K higher; the typical column density and total mass would decrease by a factor of $\sim 1.5$.  Table 3 lists total cloud masses and typical values of $N(\htwo)$, dust temperature, and \xco~measured for three different values of $\beta$.
 
\begin{figure}
\plotone{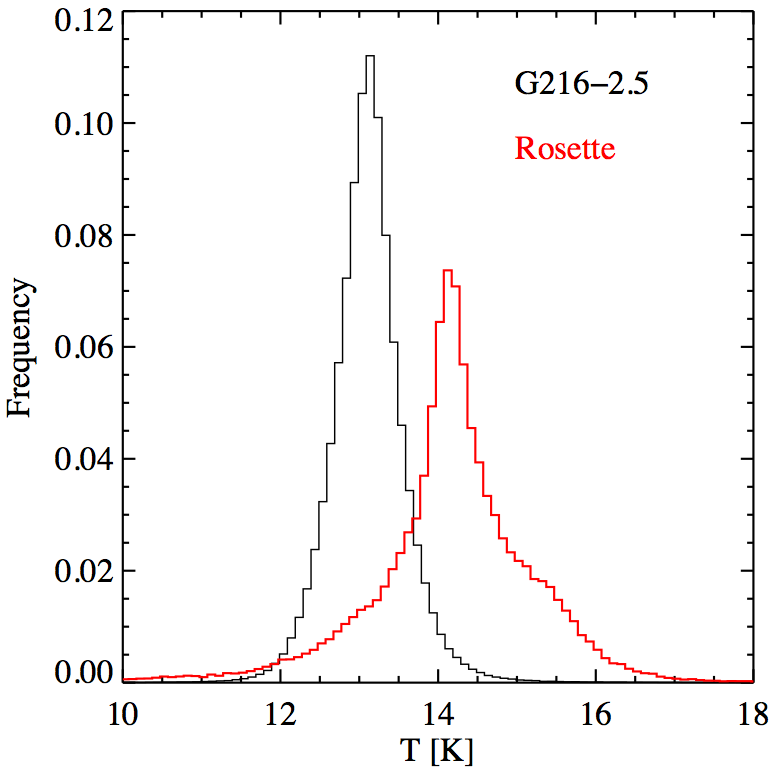}
\caption{Histograms of dust temperatures derived from \emph{Herschel} data.  The \mad~histogram (black) peaks at 13.2 K; the Rosette histogram (red) peaks at 14.2 K.  \label{fig12}}
\end{figure}

For the \mad~cloud, another source of uncertainty arises from the non-inclusion of the PACS data.  But given the low dust temperatures characteristic of \mad~($\langle T_\t{dust}\rangle=13.2~\t{K}$; Table 3) and the weakness of the PACS flux, it is not expected that the addition of the PACS data would significantly modify the results. The dust temperature histograms of the two clouds are displayed in Figure \ref{fig12}.

\begin{table}\centering
\begin{center}
\begin{tabular}{lccc}
\multicolumn{4}{c}{Table 3: Measurements for different values of $\beta$.}\\

\tableline
\textbf{\mad}                     & $\beta=2$  & $\beta=1.8$  & $\beta=1.6$   \\
\tableline
$\avg{N(\htwo)}$ [$ 10^{20}~\cm$]  & 7.8      &  5.0         & 2.8          \\
$\avg{T_{\t{dust}}}$ [K]            & 13.2     & 14.1        & 15.1      \\
$\avg{\xco}$    [$10^{20}$]        & $2.2$ & $1.4$  & $0.7$     \\
$M_{\t{MC}}$ [$10^5$ \msun]     &  1.2  & 0.78  &   0.43      \\
 
\tableline
\textbf{Rosette}               & $\beta=2$  & $\beta=1.8$  & $\beta=1.6$   \\
\tableline
$\avg{N(\htwo)}$ [$ 10^{20}~\cm$]   & 18       &  12           &   7.1        \\
$\avg{T_{\t{dust}}}$[K]   &    14.2     &   15.3             &   16.6        \\
$\avg{\xco}$  [$10^{20}$]      &$2.8$ &$1.8$  &  $1.0$  \\
$M_{\t{MC}}$ [$10^5$\msun]     &  0.36       &    0.24          &    0.15       \\
\end{tabular}
\end{center}
\end{table}

\section{Discussion}\label{sec:discussion}
This analysis has measured \xco~in two molecular clouds at sub-parsec resolutions.  The mean values are consistent with the typical $\xco\approx 2\times 10^{20}$ measured in other Milky Way MCs (e.g., Bolatto et al. 2013).  In Perseus, a molecular cloud having a similar mass to and perhaps more YSOs than the Rosette, Pineda et al. (2008) measured variations in the $X$-factor, $\xco\sim (0.9 - 1.8)\times 10^{20} $, between six subregions of the cloud.  These authors also found that the measure of \xco~is dependent upon saturation of the CO emission above $\av > 4$ mag.  In another study of Perseus, Lee et al. (2014) measured an \xco~of $\sim 3 \times 10^{19}$, an order of magnitude smaller than the mean values measured for the Rosette and \mad~in this study.  Lee et al. (2014) also  found variations across the cloud.  In their analysis of the Taurus Molecular Cloud, Pineda et al. (2010) derived $\xco\sim 2.1\times 10^{20}$ with variations in the range $(1.6-12)\times 10^{20}$.  Intriguingly, at least judging from the few studies of resolved MCs, it does not appear that \xco~depends on a cloud's evolutionary status.

\begin{figure}
\plotone{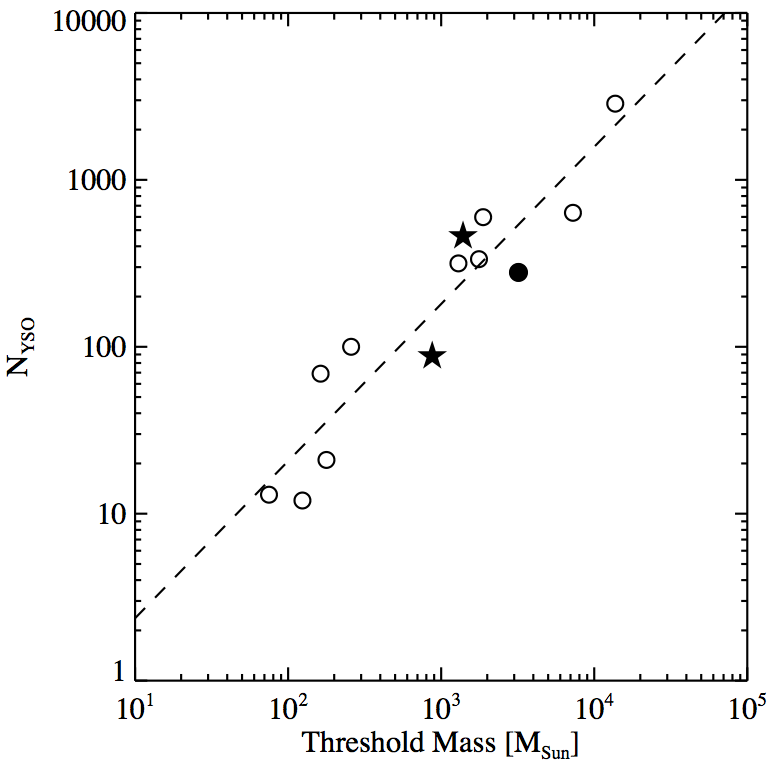}
\caption{Plot of the number of YSOs, $N_\t{YSO}$, vs. MC mass above the extinction $A_\t{K}=0.8$ mag ($\av=7.3$ mag), adapted from Lada et al. (2010).  The circles are data points taken from Lada et al. (2010), as is the best fit to the data (dashed line).  The stars represent data points for the Rosette and \mad.  The filled circle represents the California Molecular Cloud.  \label{fig13}}
\end{figure}

More insight into the evolutionary state of \mad~comes from consideration of its estimated star formation rate and its dense gas distribution.  Evidence for large variations in the star formation activity of molecular clouds has been known for some time (e.g., Mooney \& Solomon 1998), and \mad~is no longer the sole example of a massive MC with a low SFR.  In their series of extinction mapping surveys, Lada et al. (2009) identified the California Molecular Cloud, a nearby (0.5 kpc), massive ($\sim 10^5~\msun$) molecular cloud with an SFR of $\sim 70~\msun$ Myr$^{-1}$.  In a study of 11 molecular clouds from their survey, Lada et al. (2010) argued that the SFR is determined by the total amount of gas above a threshold density.  In terms of extinction, they estimated this threshold to be $A_\t{K}=0.8\pm0.2$ mag, corresponding to $\av=7.3\pm 1.8$ mag.  They compared the number of YSOs, $N_\t{YSO}$, in the clouds to their masses above the extinction threshold, $M_0$, and found a nearly linear relation of the form $N_\t{YSO}=M_0^\alpha$, with $\alpha=0.96\pm 0.11$.  From the column density maps, the total molecular mass in \mad~having extinctions above 7.3 mag is 874 \msun, about 400 \msun~less than what Lada et al. (2010) estimated for the Ophiuchus Molecular Cloud (1766 \msun).  Including all of the data available on \mad, the cloud has roughly 88 YSOs, counting the 14 identified by Lee et al. (1996) and the 74 of Megeath et al. (2009).    For the Rosette, $M_0 = 1389$ \msun, perhaps a slight underestimate of the amount of dense gas, since the entire cloud was not mapped in this study.  The known number of YSOs in this MC is $\sim 461$ (Ybarra et al. 2013).  These quantities are plotted in Figure \ref{fig13}, along with the data from the Lada et al. (2010) sample.

For the clouds in their sample,  Lada et al. (2010) assume the median mass for the stellar initial mass function (IMF) to be 0.5 \msun~(e.g., Muench et al. 2007), and they  derive an expression relating the SFR to $N_\t{YSO}$:
\begin{equation}\label{eq:yso}
\t{SFR}=0.25 N_\t{YSO}~\msun~\t{Myr}^{-1}.
\end{equation}
Using this formulation, the SFR in the Rosette is $115 \msun$ Myr$^{-1}$, roughly halfway between the rates in Taurus ($84\msun$ Myr$^{-1}$)  and Perseus ($150 \msun$ Myr$^{-1}$).  In \mad,  $\t{SFR}\approx 22 \msun$ Myr$^{-1}$, similar to the rates in the Lupus 3 and RCrA molecular clouds.  This corresponds to 44 stars Myr$^{-1}$, not far from the estimate of Megeath et al. (2009), who determined that \mad~is forming $\approx 60$ stars Myr$^{-1}$.  

Yet considering that Equation \ref{eq:yso} assumes a star formation duration of 2 Myr and that \mad~has a particularly low ratio of Class II to Class I YSOs, compared to nearby Galactic molecular clouds (Gutermuth et al. 2009; 2011), it is possible that a higher SFR in \mad~may be more true to reality.  Given the latest observations of Megeath et al. (2009) that \mad~has 33 Class I sources, which may have been born as recently as $\sim 0.5$ Myr ago, (e.g., Hatchell et al. 2007; Battersby et al. 2014),  Equation \ref{eq:yso} could be reformulated as: $\t{SFR}=0.25 N_\t{proto}~\msun~\t{Myr}^{-1} (2/t_\t{proto})$, where $t_\t{proto}$ is the typical protostar lifetime in units of Myr, and $ N_\t{proto}$ is the number of protostars.  In this case, the SFR in \mad~would be $\sim 33~\msun~\t{Myr}^{-1}$.

Lada et al. (2010) also give a predicted expression for the SFR that scales with dense mass: $\t{SFR}\approx 4.6\times 10^{-2} M_0~\msun~\t{Myr}^{-1}$, where $M_0$ is the mass of gas having $A_\t{V}\ge 7.3 \pm 1.8$ mag.  As justification, they argue that high-extinction material corresponds to material with high volume number density ($n(\htwo)\ge 10^4$ cm$^{-3}$).  Using this equation, the predicted SFR in \mad~is $\sim 40~\msun~\t{Myr}^{-1}$, about half the SFRs in Ophiuchus and Taurus, and consistent with the above estimates from Equation \ref{eq:yso}.

If anything, the predicted value may \emph{slightly} overestimate the SFR in \mad, however, due to the assumption that high-extinction material corresponds to $n(\htwo)\ge 10^4$ cm$^{-3}$.   As Clark \& Glover (2014) pointed out, the appearance of high-density gas at high extinctions does not necessarily mean high-extinction material is composed only of high-density gas.  In their simulations of clouds of different masses, Clark \& Glover (2014) found that early on in a cloud's evolution, the quantity of gas above the volume density threshold can be less than the quantity of gas above the column density threshold by orders of magnitude.  Indeed, this may be the case in \mad~if it is a young MC.  As was shown in \S \ref{sec:distribution}, \mad~has a number of high-column-density regions with low number densities and with no evidence of star formation activity.  This is in keeping with the Williams \& Blitz (1998) study of clumps in \mad~and the Rosette in the CO(3-2), CO(1-0), and CS(2-1) lines.  They found that $\sim 20\%$ of the clumps in the Rosette are star-forming or incipient star-forming and $\sim 80\%$ are ``dormant'' (i.e., non-star-forming).  In \mad, by comparison, nearly all the clumps are dormant.  Moreover, Williams \& Blitz (1998) found that dormant clumps in the Rosette tend to be denser and have steeper radial density profiles, suggesting that they are more evolved than dormant clumps in \mad.   Thus, as \mad~as a whole ages, perhaps the number of gravitationally bound cores capable of forming stars will increase, as was the case for the clouds in the Clark \& Glover (2014) simulations.  With regards to the star formation history, with the data available here and the given assumptions, the estimates of the SFR in \mad~are within only a factor of 2 of each other.  New observations in other regions of the cloud of deeply embedded sources may help to provide a more accurate measure of its star formation history.

The idea that different regions within \mad~may be evolving at different rates may also explain, in part, the regional differences in PDFs.  From Figure \ref{fig10}, the following points are of note.  First, the PDFs of star-forming regions display a power-law tail.  But at least three regions in which star formation has not been detected (6, 8, and 9) also have PDFs with an excess of high column densities (relative to log-normal), leading one to speculate that these regions might have deeply embedded sources or that star formation is incipient.  Second, there does not appear to be a characteristic width of PDFs of star-forming regions.  While signatures of star formation activity have been identified in Regions 1, 2, 5, and 7, the widths of PDFs of Regions 1 and 7 are broader than those of 2 and 5.  Moreover, the width of these latter two regions are also narrower than 3, 6, and 9, in which star formation has yet to be detected.  Third, each of the star-forming regions have PDFs which peak at higher column densities than the PDF of the cloud as a whole.  Fourth, the PDFs of regions without detected star formation---Regions 3, 4, and 6---also peak at high column densities, relative to the peak of the entire cloud (indicated by a dotted line in Figure \ref{fig10}).  Fifth, in addition to possessing high-column density gas, regions with star formation contain a significant amount of low-column-density gas, to varying degrees depending on the region.  The different sizes of the regions analyzed (Figure \ref{fig9}) do not account for these regional differences, since the characteristics of the PDFs listed above are robust to variations in the boundary sizes.  Factors, such as turbulence, that may contribute to the regional differences will be explored in a future study.

Theoretical and observational studies suggest that, as a consequence of the gravitational collapse occurring on small scales within molecular clouds, the column density PDFs evolve.  Numerical simulations of molecular cloud evolution indicate that the PDF of a cloud acquires a power-law tail or ``wing'' at high extinctions and that this tail tends to flatten over time as the star formation efficiency increases (e.g., Federrath \& Klessen 2013; Clark \& Glover 2014; Ward et al. 2014).  In their study of 23 Galactic molecular clouds, Kainulainen et al. (2009) found that actively star-forming clouds have PDFs with non-log-normal tails at high-extinctions, while quiescent, non-star-forming clouds have log-normal PDFs over the entire range of column densities.  Figures \ref{fig4} and \ref{fig5} show that both \mad~and the Rosette have high-extinction tails.  Interestingly, the slope of the \mad~PDF is slightly flatter than that of the Rosette, even when the range of values chosen for the fits are changed.  Taking this to suggest that \mad~is more evolved than the Rosette would be an over-interpretation, however, since it should be kept in mind that the slope of the high-extinction tail is probably not universal for clouds of similar age (Federrath \& Klessen 2013).  Moreover, observational column density PDFs do not easily translate to volumetric PDFs, and as discussed above, high $N(\htwo)$ does not necessarily imply high $n(\htwo)$.  To better understand (i) the regional differences in PDFs discussed above, and (2) how the properties of gas in \mad~determine its SFR, high-resolution observations of molecular lines that trace high-density gas would be useful.  With observations of $\t{NH}_3$, $\t{N}_2\t{H}$, and HCN, for instance, future work could focus on comparing cores in \mad~with those in other MCs.

In the context of observational studies that aim to understand the nature of diffuse gas in molecular clouds, the large excess over log-normal of low-column-density gas in \mad~is interesting.  Previous studies that derived column density PDFs of MCs from extinction measurements are sometimes inconclusive about the origin of this low-extinction material (e.g., Kainulainen et al. 2009; Lombardi et al. 2011). In particular, it can be difficult to ascertain whether low-extinction material is real (i.e., from the cloud of interest) or if it is mostly contamination from foreground objects along the line of sight toward the cloud.  The work presented here indicates that this low-extinction feature of the PDF persists when column densities are measured using dust emission measurements and, in the case of \mad, is likely to represent real structure. 

Spatially, the low-density tail of the \mad~PDF corresponds to low-density gas from within the cloud itself.  In the PDF represented in Figure \ref{fig4}, each column density is the excess above the boundary of the cloud, as defined by \ico~(\S 3).  Thus, the low-column density tail of the PDF does not arise from the adjacent photodissociation region, discussed in Williams \& Maddalena (1996), which lies beyond the CO boundary of the cloud.  Not including gas beyond this boundary also ensures that high-density features are not over-represented in the tail of the PDF.

It is worth considering whether this low-density gas truly arises from within \mad~itself, or is due to line of sight (LOS) contamination.  One way to test for the latter possibility is to consider how much of a contribution the low-density gas makes to the total mass of \mad.  For instance, gas having values to the left of the peak in the PDF in Figure \ref{fig4} (i.e., $N(\htwo) < 6.8\times 10^{20} \cm$) contributes to 30\% of the estimated total molecular mass.  Subtracting this contribution would result in an unrealistically low value for the total mass, so it is likely that this low-density material results from real dust emission within \mad.  Schneider et al. (2014), unlike this study, make a correction for LOS contamination toward \mad, which brings down their average column density before correction; they consider only extinctions above $\av > 1$ mag.  In this study, the contribution of atomic gas to the total hydrogen column density is corrected for (\S 4), which also has the effect of driving down the value derived here for the average column density, and values below $\av = 1$ mag are included in the analysis.  Ultimately, the combined effects of our different analyses result in Schneider et al. (2014), deriving a higher value for the peak of the \mad~PDF ($\av = 1.9$ mag, after correction) than that derived here.  

It is also instructive to consider the column density at which the PDF transitions from log-normal to the high-density power-law tail (called TP in Table 1).  In terms of visual extinction, this transition occurs at $\sim 2$ mag, in both clouds.  Kainulainen et al. (2011) called this transition extinction $A_\t{V}^\t{tail}$ and found that it occurs between 2 -- 4 mag in most of the clouds from their 2009 study, in which they derived column densities from extinction maps using star counts.  Schneider et al. (2014) derived column density PDFs for a sample of clouds, including \mad, from \emph{Herschel} dust emission maps.  For \mad, they found that  $A_\t{V}^\t{tail}\approx 4$ mag; and overall, for their group of four MCs, they found a higher range of transition extinctions occurring between 4 -- 5 mag.

In their study of cluster formation in the Rosette, Schneider et al. (2012) derived PDFs in six subregions of the cloud.  They found that the PDF for the whole cloud has a log-normal distribution up to a visual extinction of $\sim 3$ mag and a power-law tail at high extinctions ($\av=3$ -- 20 mag).  In this study, the contribution of \HI~to the total gas column density is subtracted, thus, lower values of the PDF peak and the transition extinction are derived: $A_\t{V}^\t{tail}\approx 2.1$ mag (equivalently, $N(\htwo)\approx 2.0\times 10^{20}$ \cm; see Table 1).  The essential point, nevertheless, is that the shape of the PDFs derived here---log-normal at low extinctions with power-law tails at higher extinctions---is consistent with previous studies, in spite of the different criteria for defining clouds.

Taken together, the findings of this and previous studies give the impression that \mad~is a ``young'' MC no more than a few million-years-old.  Hartmann et al. (2001) noted that the stellar populations of most molecular clouds have age spreads of no more than 1 -- 3 Myr.  This fact as well as the absence of older stars favor MCs being transient entities that live no more than $\sim 30$ Myr (e.g., Blitz \& Shu 1980).  Hartmann et al. (2001) argue for a picture in which star formation and MC formation are triggered by the same large-scale flow of distant star-forming sites.  Outflows from high-mass stars ultimately disperse the cloud of which they formed; but it is an open question as to whether low-mass stars can similarly disperse their parent cloud on short timescales.  

This line of reasoning is not necessarily at odds with the argument of Lee et al. (1994) that \mad~is a remnant cloud from past massive star formation.  One need only (re)consider the term ``remnant'' and its connotation of oldness.   If MCs form at the interface of large-scale flows, as some simulations suggest (e.g., Hartmann et al. 2001; Hennebelle et al. 2008), then new clouds naturally condense from the ``remnants'' of old clouds.  Future work related to this study will examine the dynamics of \mad~and of the atomic gas associated with it.  Williams \& Maddalena (1996) already showed that \mad~appears to be connected to S287, a distant star-forming region, by an \HI~cloud.  And Lee et al. (1994) established that \mad~has a global velocity gradient that could be explained if the MC is part of a large, expanding shell.  Another interpretation, however, is that the gradient is due to shear (if, for instance, the cloud formed at the interface of converging flows).  Considered together, this could mean \mad~recently merged from large-scale flows and has not had enough time to build up the \HI~necessary to shield \htwo~and CO from dissociating UV radiation as effectively as older MCs.  This is not a firm conclusion, nevertheless, and so subsequent work will concentrate on understanding the large scale surroundings of \mad.

\clearpage
\section{Concluding Remarks}

Although evidence for differing SFRs in MCs is not new, \mad~still seems to be exceptional, considering the peculiar combination of its properties: (1) It has a total mass comparable to the most massive MCs (e.g., California, Orion B); (2) its total dense mass is comparable to medium-sized MCs with hundreds of YSOs (e.g., Taurus, Ophiuchus); and yet (3) it forms stars at a rate similar to low-mass clouds (e.g., Lupus).  In this work, I compare the properties and distribution of molecular gas of \mad~and the Rosette, two MCs having extremely different SFRs.  I use archival \emph{Herschel} data to create high-resolution $N(\htwo)$ maps and compare these with \co{12} data to derive images of \xco~at $\sim 0.5$ pc spatial resolution for both clouds.  The main results are as follows:

\begin{enumerate}
\item For \mad,  $\avg{\xco}=2.2\times 10^{20}$, and for the Rosette, $\avg{\xco}=2.8\times 10^{20}$, values consistent with the typical \xco~measured in other Milky Way MCs (e.g., Dame et al. 2001; Bolatto et al. 2013).  The conversion factor increases to as high as $\sim 5\times 10^{20}$ toward the edges of the MCs but hovers between $\sim 0.5$ and 2 in the main bodies of the clouds.  

\item The $N(\htwo)$ PDFs of the clouds both have log-normal distributions with power-law tails at high column densities.  The \mad~PDF peaks at $N(\htwo) = 6.8\times 10^{20}~\cm$ ($\av=0.6$ mag); that of the Rosette peaks at $N(\htwo) = 1.4\times 10^{21}~\cm$ ($\av=1.4$ mag).  The Rosette has a higher fraction of its mass locked up in gas having extinctions ranging from $\av \sim 0.5$ to 15 mag.

\item To compare the properties of sites with star formation and high-density areas without stars, nine regions in \mad~are are selected for closer examination.  The PDFs of the star-forming regions and at least two of regions where star formation has not been observed appear to display power-law tails, a feature that previous authors have attributed to the dense gas associated with star formation.  Eight of the regions peak at higher extinctions than the cloud as a whole, but there is not a significant difference in the peak location between star- and non-star-forming regions.  Also, none of the regions peak close to $\av_{\t{peak}}$ measured in the Rosette. 

\item According to the \ico-\av~plots of the different regions, there is evidence for CO saturation at high extinctions in both star-forming and non-star-forming regions. Additionally, while in some regions \xco~appears to be a function of \ico, in others, \xco~is nearly constant.

\item The Rosette has a higher fraction of its mass in the form of dense gas and contains $1389\msun$ of gas above the so-called extinction threshold for star formation, $\av = 7.3$ mag.  The \mad~cloud has $874\msun$ of dense gas above this threshold.

\end{enumerate}

\hfill \break
\acknowledgments
My work is supported by the Harvard-MIT FFL Postdoctoral Fellowship.  I thank Sean Andrews, Charles Lada, and Tom Dame for their careful readings of earlier versions of this manuscript and for their helpful suggestions.   Youngung Lee and Mark Heyer have my gratitude for providing the CO data.  I thank an anonymous referee for constructive comments that improved this paper.

\end{document}